\documentclass[preprints,article,accept,pdftex,moreauthors]{Definitions/mdpi} 
\firstpage{1} 
\pubvolume{1}
\issuenum{1}
\articlenumber{0}
\pubyear{}
\copyrightyear{}
\datereceived{} 
\dateaccepted{} 
\datepublished{} 
\hreflink{} % If needed use \linebreak

\Title{
	Cosmic ray spectra and anisotropy under anisotropic propagation model with spiral galactic sources
}

\TitleCitation{Title}

\Author{Aifeng Li $^{1,}$\orcidA{}, Zhaodong Lv $^{1}$,	 Wei Liu $^{2,3,4,}$*  , Yiqing Guo $^{2,3,4,}$*  and   Fangheng Zhang $^{1}$}

\address{%
$^{1}$ \quad College of Information Science and Engineering, Shandong Agricultural University, Taian 271018, China;

$^{2}$ \quad Key Laboratory of Particle Astrophysics, Institute of High Energy Physics, Chinese Academy of Sciences, Beijing 100049, China;

$^{3}$ \quad University of Chinese Academy of Sciences, Beijing 100049, China;
	
$^{4}$ \quad TIANFU Cosmic Ray Research Center, Chengdu 610000, China;
\\

}

\corres{Correspondence: liuwei@ihep.ac.cn (W.L.) and guoyq@ihep.ac.cn {(Y.Q.)}}

\abstract{
In our previous work, we have investigated Galactic cosmic ray (GCR) spectra and anisotropy from 100 GeV to PeV, under anisotropic propagation model with axisymmetric  distributed galactic sources.
Numerous  observational  evidence   have  indicated  that  the  Milky
Way is a typical spiral galaxy. 
In this work, we further utilize anisotropic propagation models with  spiral galactic sources to investigate  spectra and anisotropy of CRs.
During the calculation process, we employ the spatially dependent diffusion (SDP) model with different diffusion coefficients for the inner and outer halo, while the background CR sources  is spiral distribution.
To better explain the anomalous observations of nuclear spectral hardening at $ {\cal R}\sim$ 200 GV and the complicated energy dependence of anisotropy from GeV to PeV, we introduce the contribution of the  nearby Geminga source. 
Additionally, we incorporate the impact of the local regular magnetic field (LRMF) and the corresponding anisotropic diffusion on large-scale anisotropy within the SDP model. 
By comparing the spiral and axisymmetric distribution models of background sources, it is found that both of them can well reproduce the CR spectra and anisotropy from 100 GeV -PeV.
However, there exist differences in their propagation parameters.
The diffusion coefficient with  spiral distribution of sources is larger than that with axisymmetric distribution,  and its  spectral indices are slightly harder.
Future high-precision measurements of CR anisotropy, such as LHAASO experiment, will be crucial in evaluating the validity of our proposed model.
}

\keyword{Galatic cosmic rays; Spiral arm structure; Cosmic ray anisotropy; Cosmic ray spectra} 

\begin{document}
%\linenumbers 

%%%%%%%%%%%%%%%%%%%%%%%%%%%%%%%%%%%%%%%%%%
\setcounter{section}{-1} %% Remove this when starting to work on the template.

\section{Introduction}

With the improvement of CR observation technology, the new generation of  experiments have entered the era of high precision measurement and unveiled a series of unexpected phenomena.
In recent years, multiple experiments such as  ATIC-2\cite{ref-journa1} 
%\cite{2007BRASP..71..494P,20=09BRASP..73..564P}%, 
CREAM\cite{ref-journa2,ref-journa3} PAMELA
\cite{ref-journa4} and AMS-02 \cite{ref-journa5,ref-journa6} DAMPLE\cite{ref-journa7}and calorimeter experiment CALET \cite{ref-journa8} have observed that the spectra of protons and helium nuclei become harder at ${\cal R}$$\sim$$200 $ GV. Furthermore, DAMPE\cite{ref-journa9}, CREAM\cite{ref-journa10} and NUCLEON\cite{ref-journa11}  found  spectra become soften at ${\cal R}$$\sim$$14 $  TV. 
%It is clear that this anomaly deviates significantly from the expected power-law energy spectrum of cosmic rays.
This subtle anomaly of  spectra obviously deviates from the expected CR power law spectrum, and has emerged as a focal point of theoretical research in recent years.
The main theoretical explanations for the anomaly  are:
the nearby sources near the solar system contribute to the "bulge" of the CR spectra\cite{ref-journa12,ref-journa13};  interaction between CRs and accelerating shock waves\cite{ref-journa14,ref-journa15};  CR propagation process effect\cite{ref-journa13,ref-journa16};  multiple acceleration sources superimposed factors\cite{ref-journa17,ref-journa18}.

CRs, mostly charged particles, become isotropic as they travel through the Milky Way due to deflection by the Galactic magnetic field (GMF).
However, subtle CR anisotropy with relative amplitudes in the order of $10^{-4} \sim 10^{-3}$ is  observed  at a wide energy range from 100 GeV to PeV by a large number  of underground $\mu$ detectors and EAS  array experiments.
Tibet \cite{ref-journa19,  ref-journa20,  ref-journa21},Super-Kamiokande \cite{ref-journa22}, Milagro \cite{ref-journa23, ref-journa24}, IceCube/Ice-Top \cite{ ref-journa25, ref-journa26,ref-journa27,ref-journa28,ref-journa29}, ARGO-YBJ \cite{ref-journa30, ref-journa31}, EASTOP\cite{ref-journa32},
KASCADE\cite{ref-journa33,ref-journa34} HWAC\cite{ref-journa35,ref-journa36} have revealed the complex  evolution  of anisotropy with energy.
Experimental results show that the amplitude of anisotropy  increases first and then decreases with energy below 100 TeV,  but gradually increases again above 100 TeV. At the same time the phase is reversed at about 100 TeV. It is clear that both amplitude and phase contradict  the expectations of the conventional propagation model.
In general, the origin of anisotropy may consist of  the following reasons: nearby sources near the solar system\cite{ref-journa13,ref-journa37}, the deflection of local regular magnetic  field\cite{ref-journa37,ref-journa38,ref-journa39}, CR propagation\cite{ref-journa13}
and Compton-Getting effect caused by the relative motion between Earth rotation and CRs\cite{ref-journa40}.

CR spectra and anisotropy from  GeV to $\sim 100 $ TeV have some common anomalous characteristics, suggesting that they may have a common origin.
In our previous work\cite{ref-journa37,ref-journa41}, based on the assumption that CR  background sources follow the axisymmetric distribution, we used the SDP model 
to calculate the  CR spectra and anisotropy. In SDP model, we innovatively introduce the significant contribution of nearby sources and the anisotropic diffusion effect of LRMF on CR particles.  The hybrid model successfully reproduces the fine structure of the nuclear spectra and the complex characteristics of the anisotropy with energy.
The current extensive experimental observations clearly reveal that the Milky Way is a typical spiral galaxy.  The spiral arms, where high density gas accumulates, are hotspots for rapid star formation\cite{ref-journa42,ref-journa43}. Therefore, there is a high correlation between the distribution pattern of CR sources (especially supernova remnants SNRs) and the spiral arm structure.
Until recent years, the impact of the spiral distribution of CR sources has garnered attention in research. Several studies  have demonstrated that this spiral distribution of CR sources can significantly influence the positron and electron spectra, which offer a more compelling explanation for the observed excess of positron and electron at 30 GeV\cite{ref-journa44,ref-journa45}.
Does the spiral distribution of CR background sources  affect CR anisotropy?  %Currently, under the SDP model, the anisotropy associated with a spiral distribution of background sources has not yet been studied.
This work aims to analyze the CR spectra and anisotropy by applying SDP model with spiral distribution of CR background sources. 
%During the computational process, we also take into account the contributions from the Geminga nearby source and anisotropic diffusion effects on CR particles induced by the LRMF.

% with different diffusion coefficients of inner and outer halo

The rest of the paper is organized as follows: Section \ref{sec: Model} introduces the model in detail, including the SDP model, the spiral structure of background sources, nearby sources, anisotropic diffusion and large-scale anisotropy. In Section \ref{sec:results}, the results of CR spectra and anisotropy are presented and thoroughly discussed; Section \ref{sec:summary} offers the summary.

%In recent years, a large number of studies have supported  that nearby sources are well correlated with the anomalies.
% the association of nearby source with these anomalies. 
%nearby sources are well correlated with the anomalies. % of CR spectra and anisotropy.
%Which nearby source contributes to the anomalies of spectra and anisotropy is an important research topic. 
%\cite{ref-journa38} shows that the Geminga source and anisitorpic diffusion of CRs induced by the LIMF  can explain both nuclear spectra and anisotropy.
%\cite{ref-journa43} demonstrates that three local SNRs, i.e., Geminga, Monogem, and Vela, could have important contributions to both proton and electron spectra. However, the expected anisotropy from Monogem is obviously inconsistent with the observations. And  the bump in electron spectrum above several TeV could stem from the young Vela SNR.
%\cite{ref-journa44} demonstrates that  only Geminga SNR could be the proper candidate of the local CR source by fitting calculation.
%\cite{ref-journa45} have found that Monogem can reasonably account for primary electron excess and proton spectrum. 
%\cite{ref-journa40} presents that an excellent candidate of the local CR source responsible for the dipole anisotropy at $ 1\sim 100$ TeV is the Vela SNR.

%In this manuscript we will briefly illustrate the construction of our model and discuss its Galactocentric radial distribution.

%%%%%%%%%%%%%%%%%%%%%%%%%%%%%%%%%%%%%%%%%%
\section{Model Description}\label{sec: Model}

\subsection{Spatially dependent diffusion}

In recent years, the SDP model of CRs has been proposed and widely applied. It was initially introduced as a two-halo model  to accoount for  the excess of primary proton and helium fluxes at $\cal R\sim$ 200 GV\cite{ref-journa46}. 
Afterwards, it was further used to explain the excess of secondary and heavier components \cite{ref-journa44,ref-journa47,ref-journa48,ref-journa49},
 diffuse gamma-ray distribution \cite{ref-journa50},
 and large-scale anisotropy \cite{ref-journa51,ref-journa52,ref-journa53}. 
The recent observation of  halos around pulsars has revealed  that  CRs diffuse much slower than that inferred from B/C ratio,
 which strongly supports the assumption that diffusion could be spatially dependent\cite{ref-journa54,ref-journa55}.

%HAWC experiment obtained by observing B/C that the diffusion coefficient of CRs near the source of the galactic disk was two orders of magnitude lower than that away from the source.
%In recent years, the SDP model of CRs has been proposed and widely applied\cite  {ref-journa17,ref-journa52,ref-journa53}. %with different propagation coefficients in the near and away galactic disk
%has been proposed and widely applied\cite {ref-journa17,ref-journa52,ref-journa53}.
%the inner halo and outer halo has been proposed and widely applied.
%In the SDP model, the  galactic diffusion halo is divided into two regions, i.e  inner halo (IH) and outer halo (OH). 
In the SDP model, the galactic diffusive halo is delineated into two distinct zones: inner halo (IH) and outer halo (OH). 
The galactic disk and its surrounding area are referred to as the IH, while the extensive diffusive region outside IH is called OH.
In IH region, where are more  sources, the activity of supernova explosion  will lead to more intense turbulence. Consequently, the diffusion of CRs in IH region is slowed down, and the diffusion coefficient exhibits a lesser dependence on rigidity.
Whereas in OH region, the diffusion of CRs is less affected by stellar activity, so CRs diffuse faster. And diffusion coefficient is consistent with the conventional propagation model and only depends on rigidity.
%The galactic disk, along with its immediate vicinity, comprises the IH, whereas the region extending beyond the IH constitutes the OH.
%The galactic disk and its surrounding region is called the IH,  while the diffusion region outside the IH is called the OH. 
%The Galactic disk and its surrounding area are called the IH region,
%Within the IH, characterized by a higher concentration of sources, the frequent occurrence of supernova explosions generates more intense turbulence.
%In IH region, where are more  sources, the activity of supernova explosion  will lead to more intense turbulence. Consequently, the diffusion of CRs in this region is slowed down, and the diffusion coefficient exhibits a lesser dependence on rigidity.
%The outer diffuse region of the IH is referred to as the OH.
%Whereas in OH region, the diffusion of CRs is less affected by stellar activity, and diffusion coefficient is consistent with the traditional propagation model and only depends on rigidity.

In this work, we adopt  SDP model and the diffusion coefficient  is parameterized as\cite {ref-journa37, ref-journa41}
\begin{linenomath*}
\begin{align}
\ D_{xx}( r,z,\cal R) &=
D_0F(r,z)({\dfrac{\cal R}{ {\cal R}_0}})^{{\delta_0}F(r,z)}
\label{eq:D_{xx}}
\end{align}
\end{linenomath*}

where $r$ and $z$ are cylindrical coordinates, $\cal R$ is particle's rigidity and $D_0$ is a constant. 
The reference rigidity $\cal R_0$ is fixed to 4 GV.
 The parameterization of $F(r,z)$ can be parameterized as
\begin{linenomath*}
\begin{align}
\begin{split}
F(r,z)= \left \{
\begin{array}{ll}
g(r,z)+[1-g(r,z)]{(\dfrac{z}{ \xi z_0})}^n,       &  |z|\leq\xi z_0 \\
1, & |z|>\xi z_0\\
\end{array}
\right.
\end{split}
\end{align}
\end{linenomath*}
 where $g(r,z) = N_m/[1+f(r,z)]$, and $f(r,z)$ is the source density
distribution.   The total half thickness of the propagation halo is $z_0$, and  IH and OH are  $\xi z_0$ and  $(1-\xi) z_0$ respectively.

Some propagation codes can simulate the process of CR propagation, for example: GALPROP \cite{ref-journa56},  DRAGON \cite{ref-journa57} and  PICARD \cite{ref-journa58}. In this work, we adopt numerical package DRAGON to solve the CR transport equation .

\subsection{Spiral distribution of CR sources}

In view of  the  diffusion  length  of  CRs  is  usually  much  longer 
than the characteristic spacing between the adjacent spiral  arms, CR source are generally approximated as axisymmetric and parameterized to
\begin{linenomath*}
	\begin{equation}
	f(r,z) = (r/r_\odot)^{\alpha }
	\exp[-\beta (r -r_\odot)/r_\odot] \exp(-|z|/z_s),
	\label{con:SNRs}
	\end{equation}
	
\end{linenomath*}
where $r_\odot = 8.5$ kpc represents  the  solar  distance  to  the  Galactic  Center (GC).
The parameters $\alpha$ and  $\beta$ are taken as 1.69 and 3.33 respectively\cite{ref-journa59}. Perpendicular  to  the  Galactic  plane,  the  density  of  CR
sources descends as an exponential function, with a mean  value $z_s = 0.2$ kpc.

However, a large number of observations have indicated that the Milky Way is a typical spiral galaxy\cite{ref-journa42, ref-journa43}.
The spiral arms where high-density gas accumulates are regions of rapid star formation.
In order to more accurately describe the source distribution, the  CR source $f(r,z)$ adopts  spiral distribution.
%the CR sources are approximated as spiral distributed in this work.
In this work, a model established by Faucher-Giguere Kaspi was used to describe the spiral distribution \cite{ref-journa60}.
The Galaxy consists of the four major  spiral arms extending outward from the galactic center: Norma, Carina-Sagittarius, Perseus, and Crux Scutum. 
And the locus of the i-th arm centroid expressed as a logarithmic curve: $\theta(r)=k^i\ln(r/r_0^i)+\theta_0^i$, where $r$ is the distance to the GC. The values of $k^i$, $r_0^i$ and $\theta_0^i$  for each arm refer to \cite{ref-journa44}. Along each spiral arm, there is a spread in the normal direction which follows a Gaussian distribution, i.e.
\begin{linenomath*}
	\begin{equation}
	f_i=   {\dfrac{1}{ \sqrt{\pi2}\sigma_i}}
	\exp[-\dfrac{(r-r_i)^2}{2\sigma_i^2}, i\in[1,2,3,4]
	\end{equation}
\end{linenomath*}
where $r_i$ is the inverse function of the $i$-th spiral arm’s locus and the standard deviation $\sigma_i^2$ is $0.07 r_i$.
The number density of SNRs at different radii is still consistent  with  the  radial  distribution  in  the  axisymmetric case, i.e. equation \eqref{con:SNRs}. % Eq (3).

The injection spectrum of background sources is assumed to be a power-law
of rigidity with a high-energy exponential cutoff, $q({\cal R}) \propto
{\cal R}^{-\nu} \exp(-{\cal R}/{\cal R}_{\rm c})$. The cutoff rigidity of each element could be either $Z$- or $A$-dependent.

\subsection{Nearby source}

In this work, we adopt the Green's function method to solve the time-varying propagation equation of CRs from nearby sources under spherical geometry conditions, assuming infinite boundary conditions\cite {ref-journa61,ref-journa62}.

As for the instantaneous and point-like injection, the CR density of a nearby source as a function of location, time, and rigidity, is computed using 
\begin{linenomath*}
\begin{equation}
\phi(r,{\cal R},t)=\frac{q_{\rm inj}({\cal R})}{(\sqrt{2\pi}\sigma)^3}
\exp\left(-\frac{(r-r')^2}{2\sigma^2}\right),
\end{equation}
\end{linenomath*}
where $q_{\rm inj}({\cal R})$ is parameterized
as a power-law function of rigidity with an exponential cutoff.i.e, $q_{\rm inj}({\cal R})=q_0{\cal R}^{-\alpha}
\exp(-{\cal R}/{\cal R}'_{\rm c})$, $\sigma({\cal R},t)=\sqrt{2D({\cal R})t}$ 
is the effective diffusion length within time $t$, $D({\cal R})$ is the 
diffusion coefficient which is adopted as the value nearby the solar system. 

In our previous works, the spectral anomaly at  200 GeV and dipole anisotropy below 100 TeV are attributed to the Geminga SNR. The Geminga SNR is located in the direction of $l=194.3^\circ, b=-13^\circ$ and its distance to the solar system is $d \sim 330 pc$\cite {ref-journa63}.  Its explosion time was about $\tau =3.4\times10^5$ years ago, which is inferred from the spin-down
luminosity of the Geminga pulsar\cite {ref-journa64}.
In this work, we also select the Geminga SNR as the optimal source.

%%%%%%%%%%%%%%%%%%%%%%%%%%%%%%%%%%%%%%%%%%
%\section{  Anisotropic diffusion and Large-scale anisotropy}\label{sec: Anisotropic diffusion}
\subsection{ Anisotropic Diffusion and Large-Scale Anisotropy}
%In the standard propagation model,  propagation of CRs is isotropic. The amplitude of the dipole anisotropy is proportional to the spatial gradient of the CR density and the diffusion coefficient, which  the latter is a scalar that depends solely on rigidity. The anisotropy can be written as \cite{ref-journa40, ref-journa60}
%\begin{linenomath*}
%\begin{equation}
%{\delta} = \dfrac{3D}{v}  \dfrac{\nabla\psi}{\psi} ~.
%\label{eq:Anisotropy_iso}
%\end{equation}
%\end{linenomath*}

By observing neutral particles passing through the heliosphere boundary,  the IBEX experiment has unveiled that the LRMF aligns with coordinates $(l,b = 210.5^\circ,-57.1^\circ)$, within a $20$ pc radius\cite{ref-journa65}.  We have discovered that the direction of the LRMF  is generally consistent with the CR anisotropy observed  below 100 TeV. 
Some research has also revealed  that the TeV cosmic ray anisotropy is associated with the LRMF\cite{ref-journa37, ref-journa38, ref-journa39, ref-journa41}.

When CRs are deflected by magnetic fields, they diffuse anisotropically.
It is generally believed that CRs diffuse faster along the direction of the magnetic field than perpendicular to it.  The corresponding dipole anisotropy is expected
to be modified by LRMF.  In this scenario, the diffusion coefficient $D$ is replace of a tensor $D_{ij}$.
The $D_{ij}$ associated with the magnetic field is written as
\begin{linenomath*}
\begin{equation}
D_{ij}\,\equiv\,D_\perp\delta_{ij}\,+\,\big(D_\|-D_\perp\big)b_ib_j ~, ~ ~  
b_i = \dfrac{B_i}{|\vec{B}|}
\label{eq:D_ij_1}
\end{equation}
\end{linenomath*}
Where $D_{\parallel}$ and $D_{\perp}$ are the diffusion coefficients aligned parallel
and perpendicular to the ordered magnetic field, $b_i$ is the $i$-th component of the unit vector  \cite {ref-journa63},  respectively.
The values  $D_\parallel$ and $D_\perp$   is parameterized as a power-law function of rigidity,\cite {ref-journa41, ref-journa66}
%of rigidity in this work refer to the form\cite {ref-journa38,ref-journa64} and are shown as follows, 
\begin{linenomath*}
\begin{align}
D_\parallel &= D_{0\parallel} \left(\frac{\cal R}{{\cal R}_0} \right)^{\delta_\|} ~, \\
D_\perp &=\,D_{0\perp} \left(\frac{\cal R}{{\cal R}_0} \right)^{\delta_\perp} \equiv \varepsilon D_{0\parallel} \left(\frac{\cal R}{{\cal R}_0} \right)^{\delta_\perp} ~,
\label{eq:DparaDperp}
\end{align}
\end{linenomath*}
where $\varepsilon = \dfrac{D_{0\perp}}{D_{0\parallel} }$ is the ratio between perpendicular and parallel diffusion coefficient at the reference rigidity ${\cal R}_{0}$.

Under the anisotropic diffusion model, the dipole anisotropy can be written as,
\begin{linenomath*}
\begin{equation}
{\delta} = \dfrac{3D}{v}  \dfrac{\nabla\psi}{\psi}= \dfrac{3}{v \psi} D_{ij}  \dfrac{ \partial \psi}{ \partial x_j} ~.
\label{eq:Anisotropy_ani}
\end{equation}
\end{linenomath*}

%%%%%%%%%%%%%%%%%%%%%%%%%%%%%%%%%%%%%%%%%%
\section{Results and Discussion}\label{sec:results}

\subsection{Proton and Helium spectra of nearby sources}

Since the spatial scale of the LRMF is significantly smaller than the average propagation length of CRs deduced from the B/C ratio, LRMF doesn't have remarkable  impact on the energy spectra. Therefore, in this work, the isotropic diffusion SDP model is used to calculate the nucleon energy spectra.

\begin{figure}[htb]
	\centering
	\includegraphics[width=0.68\textwidth]{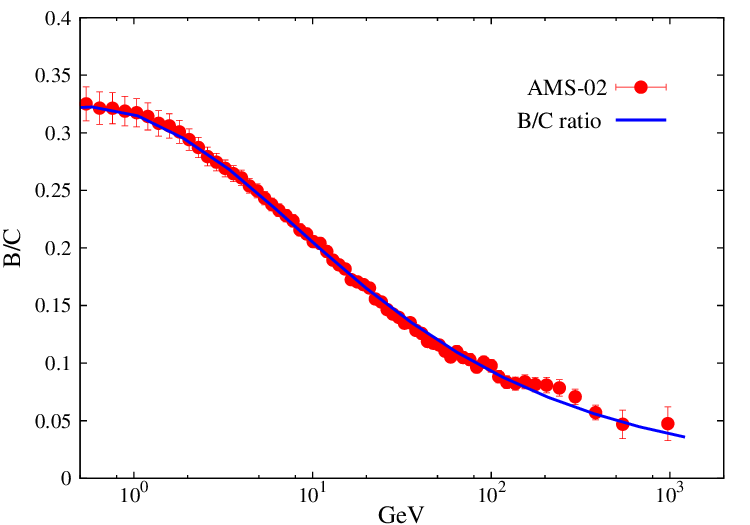}
	\caption{Fitting to B/C ratio with the model prediction.
		The B/C data points are taken from AMS-02 experiment \cite{ref-journa67}. % (Aguilar~et~al. 2016).
	}
	\label{fig:BCratio}
\end{figure}

Firstly, the propagation parameters for the SDP model  can be determined  by fitting the B/C ratio. % and $^{10}\!B/^{9}\!B$ ratios.
Figure \ref {fig:BCratio} presents the fitting results of the B/C %and $^{10}\!B/^{9}\!B$ 
ratio, which align well with the experimental data from AMS-02\cite{ref-journa67}.
%The propagation parameters are summarized in Table 1.
The corresponding propagation  parameters are respectively $D_0=9.0 \times 10^{28}{\rm cm}^2$,  ${{\cal R}_0}= 4 $ GV, $ \delta_0=0.58$, $N_{\rm m}=0.5$, $\xi=0.12$.  The Alfvénic velocity is $v_A= 6 {\rm km}\cdot{\rm s}^{-1}$, and thhe half thickness of the propagation halo is  $z_0=5 \rm kpc$.

The normalization, power index, and cutoff rigidity  for each element of  the background source injection spectra are obtained by fitting with  energy spectra of experimental observations. 
The cutoff rigidities of different compositions are regarded as the limits of acceleration in the sources and assumed to be Z dependent with high-energy exponential cutoff.
Similarly, the injection spectra of the nearby source is also set using the same method.

\begin{table}[H]
	\small
	\caption{Injection parameters of the background and nearby Geminga source.}
	\label{tab:para_bk_inj}
	%\begin{center}
	\begin{adjustwidth}{}{}%{-\extralength}{5cm}
		
		%\centering %% If there is a figure in wide page, please release command \centering
	
			\newcolumntype{C}{>{\centering\arraybackslash}X}
		\begin{tabular}{m{1.0cm}<{\centering}m{3.4cm}<{\centering}m{0.9cm}<{\centering}m{1cm}<{\centering}m{2.3cm}<{\centering}m{1cm}<{\centering}m{1cm}<{\centering}m{0.1cm}} 
			\toprule
			& \multicolumn{3}{c}{\textbf{Background}} & \multicolumn{3}{c}{\textbf{Geminga  Source}}
			 \\
			\midrule
			%\toprule[1.5pt]
			\textbf{Element }& \textbf{Normalization} \boldmath{$^\dagger$} & \boldmath{$\nu$}& \boldmath{$\mathcal R_{c}$} & \boldmath{$q_0$}& \boldmath{$\alpha$} & \boldmath{${\cal R}'_c$} & \\
			\midrule
			& \boldmath{$({\rm m}^2 {\rm sr} {\rm s}{\rm GeV})^{-1}$} & & \textbf{PV} & \textbf{GeV}\boldmath{$^{-1}$} & &  \textbf{TV}  \\
			\midrule
				p   & $4.36\times 10^{-2}$    & 2.30  &  5 & $7.74\times 10^{52}$  & 2.16 & 22\\
			He & $2.27\times 10^{-3}$   & 2.21    &  5  & $2.35\times 10^{52}$  & 2.10  &  22\\
			C   & $1.0\times 10^{-4}$   & 2.24    &  5  & $7.80\times 10^{50}$    & 2.13 &  22  \\
			N   & $1.16\times 10^{-5}$   & 2.20    &  5  & $1.03\times 10^{50}$  & 2.13 &   22  \\
			O   & $1.24\times 10^{-4}$   & 2.25    &  5  & $9.0\times 10^{50}$ & 2.13  &   22  \\
			Ne & $1.22\times 10^{-5}$   & 2.20   &  5  & $1.10\times 10^{50}$ & 2.13  &  22 \\
			Mg & $1.83\times 10^{-5}$   & 2.23   &  5  & $1.02\times 10^{50}$ & 2.13  &  22 \\
			Si & $2.35\times 10^{-5}$     & 2.29  &  5  & $1.02\times 10^{50}$ & 2.13  &   22\\
			Fe & $2.47\times 10^{-5}$    & 2.26    &  5  & $2.75\times 10^{50}$ & 2.13   &  22\\
			\bottomrule
			%\bottomrule[1.5pt]
		\end{tabular}
		
	\end{adjustwidth}
	\footnotesize{$^\dagger$ {The normalization is set at total energy $E = 100$ GeV.}}
\end{table}
\unskip
\begin{figure*}[!ht]
	\includegraphics[width=0.52\textwidth]{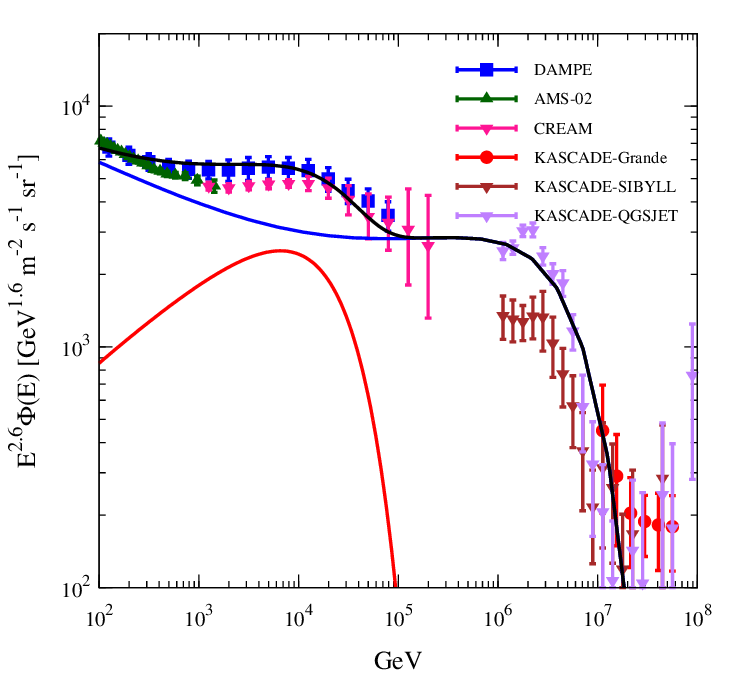}
	\includegraphics[width=0.52\textwidth]{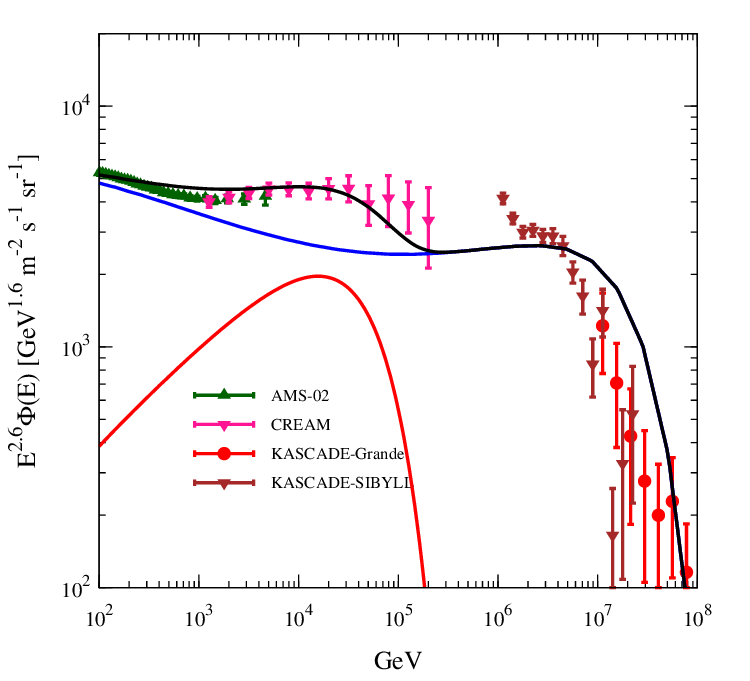}
	\caption{Energy spectra of protons (left) and helium nuclei (right). 
		The blue and red lines are the background fluxes and the fluxes from a nearby Geminga SNR source, respectively. The black lines represent the total fluxes.
		The data  points are taken from 
		DAMPE\cite {ref-journa7, ref-journa9},
		AMS-02 \cite{ref-journa5, ref-journa6},
		CREAM-III \cite{ref-journa10}, NUCLEON \cite{ref-journa68}, KASCADE \cite{ref-journa69} and KASCADE-Grande \cite{ref-journa70} respectively. 
		}
	\label{fig:Energy spectra}
\end{figure*}

\begin{figure*}
	\centering
	\includegraphics[width=0.7\textwidth]{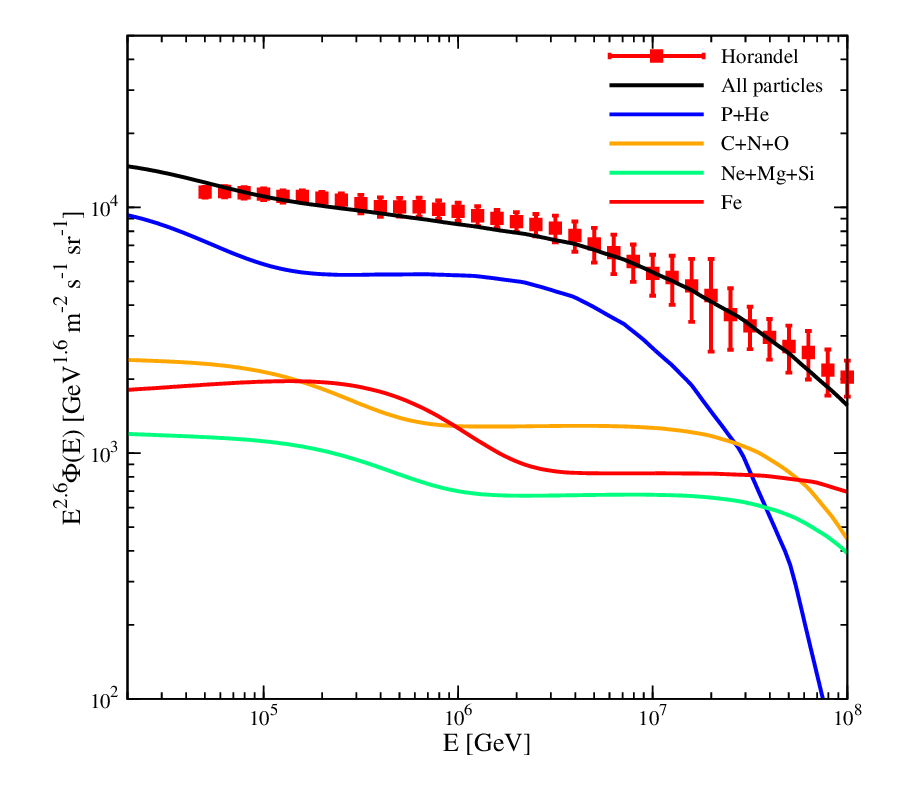}
	\caption{The all-particle spectra multiplied by $E^{2.6}$.
		The data points are taken from ref  \cite{ref-journa71}.
		The solid lines with different colors are the model predictions of different mass groups, and the black solid line is the total contribution.
	}
	\label{fig:all_spec}
\end{figure*}

Figure  \ref{fig:Energy spectra}  shows the proton (left) and helium (right) spectra, in which red, blue, and black lines are the contributions from nearby Geminga  source, background sources, and sum
of all, respectively. The corresponding injection parameters of different nuclei in the background and nearby sources are shown in Table~\ref{tab:para_bk_inj}.
The spectral indices of the nearby source component are assumed to be slightly harder than that of the background component,  which helps fit the data better.
In order to explain the softening observed at tens of TeV in the proton and helium spectra, the cutoff rigidity of the local source has been determined to be 22 TV. Additionally, to accurately depict the all-particle spectrum and the cutoffs of proton and helium at PeV energies, the cutoff rigidity of the background sources is set at 5 PV.
It can be seen that the contribution by the nearby Geminga  source can simultaneously account for both the spectral hardening features
at $\mathcal R \sim 200$ GV and the softening features at $\mathcal R \sim 10$ TV.
We have also presented the results of the all-particle spectrum of CRs, as shown in Figure \ref{fig:all_spec}. The results   are in good agreement with the Horandel experiment and  successfully reproduce the "knee" structure.

\subsection{Anisotropy}
\begin{figure*}
	\includegraphics[width=1.\textwidth]{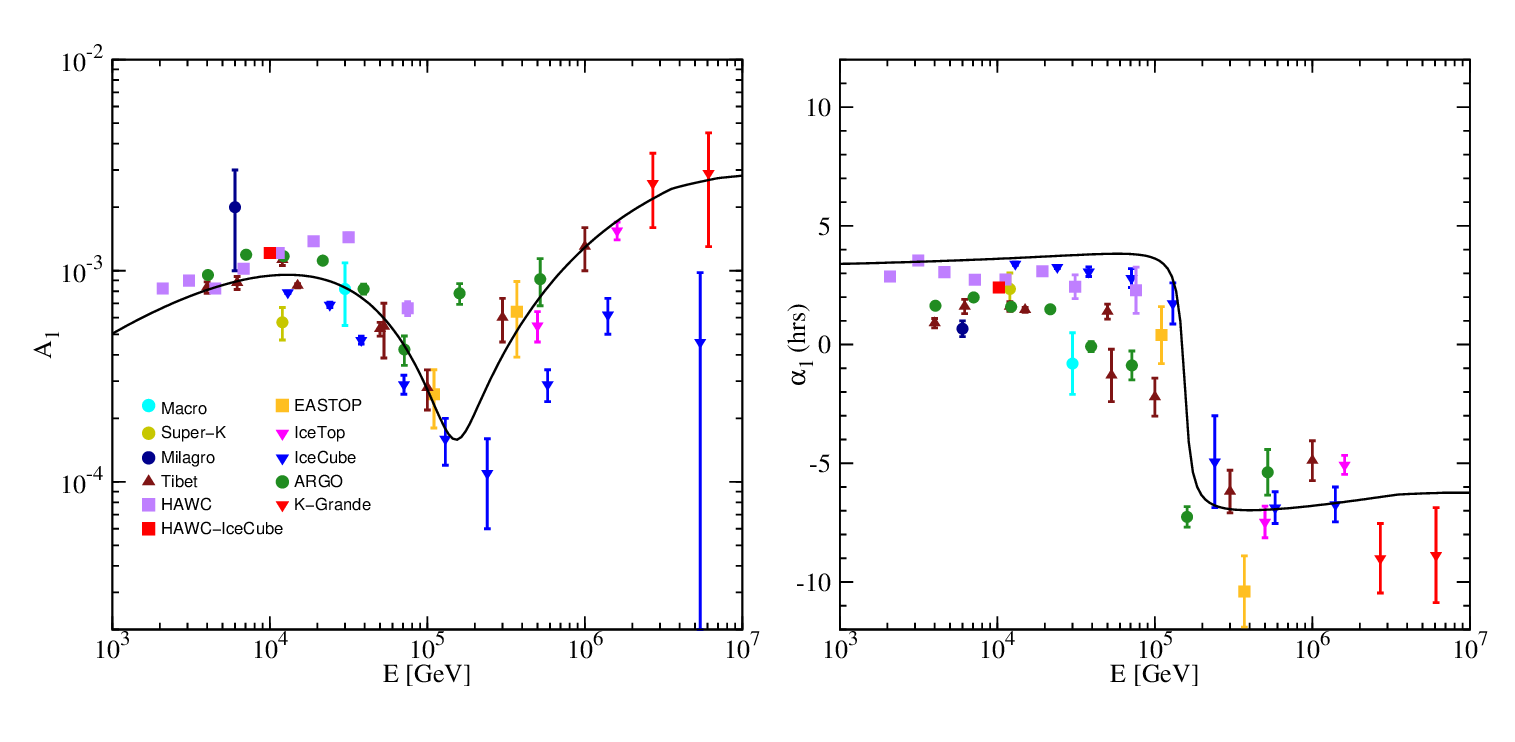}
	\caption{The amplitude (left) and phase (right)	of  anisotropy with the contribution from nearby Geminga  source. The~data points are taken from 
		Marco~\cite{ref-journa72},
		Super-Kamiokande~\cite{ref-journa22},
		EAS-TOP~\cite{ref-journa32,ref-journa73},
		Milagro~\cite{ref-journa24},
		IceCube~\cite{ref-journa25,ref-journa27,ref-journa29},
		Ice-Top~\cite{ref-journa28},
		ARGO-YBJ~\cite{ref-journa31},
		Tibet~\cite{ref-journa20, ref-journa21, ref-journa74},
		KASCADE-Grande~\cite{ref-journa33,ref-journa34}
		HAWC~\cite{ref-journa39}, and
		HAWC-IceCube~\cite{ref-journa39}.
	}
	\label{fig:ani_spiral}
\end{figure*}

Unlike the energy spectra of CRs, the LRMF can dsignificantly deflect the propagation direction of CRs, thereby influencing the dipole anisotropy. Therefore, in the process of calculating the anisotropy, we introduce the anisotropic diffusion effect of CRs induced by the LRMF.
The parameters of parallel diffusion coefficient $D_\parallel$ are set as those in \textsl {Section $1.1$}.
CRs from TeV to  PeV energy region are thought to travel faster parallel to the magnetic field than perpendicular to it, therefore  we set $D_\parallel> D_\perp$,  $\varepsilon=0.01$ and the difference between  $\delta_\perp$ and $\delta_\parallel$ is 0.32.

Figure  \ref{fig:ani_spiral}  presents the evolution of  both amplitude and phase of anisotropy with the energy,  incorporating contributions from Geimga SNR and LRMF, within the context of the spiral distribution of background sources. %of the spiral arm distribution.
It is obvious that both phase and amplitude agree well with experimental data, which  validates the reasonability of our model.
Below 100 TeV, the  phase  points in the direction of the LRMF.  The results indicate that Geminga source and the deflection of LRMF dominate the anisotropic phase, although the nearby flux is sub-dominant.
Above 100 TeV, the phase points to GC indicates  background sources dominate, since galactic CR sources are more abundant in the inner galaxy.

We compared the results of  both spiral  and  axisymmetric distribution  of  background  sources, which correspond to the results of this work and previous work\cite{ref-journa37, ref-journa41}. It was found that   the calculated energy spectra and anisotropies from both background source distribution models are in agreement with the experimental results well. However, their propagation parameters are different, which is attributed to the fact that the source distribution affects the propagation of CRs.
The diffusion coefficient with  spiral distribution of sources is larger than that with axisymmetric distribution,  and its  spectral indices are slightly harder.

\begin{figure}[H]
\begin{adjustwidth}{-\extralength}{0cm}
	\includegraphics[width=0.45\textwidth]{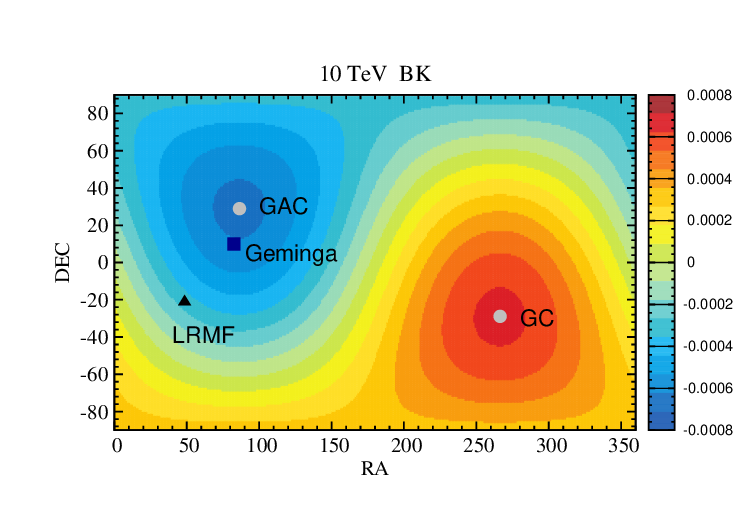}
	\includegraphics[width=0.45\textwidth]{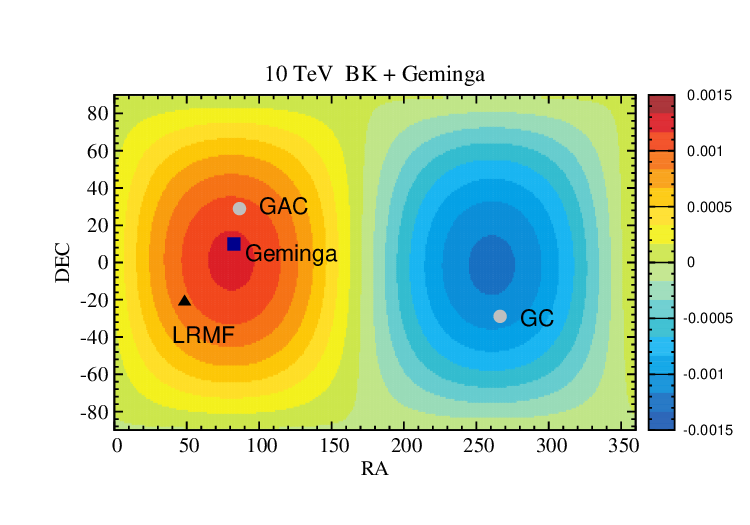}
	\includegraphics[width=0.45\textwidth]{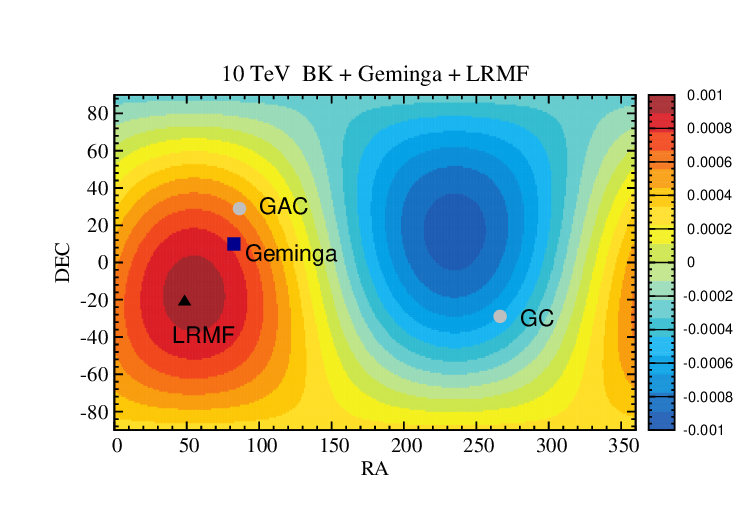}
	\includegraphics[width=0.45\textwidth]{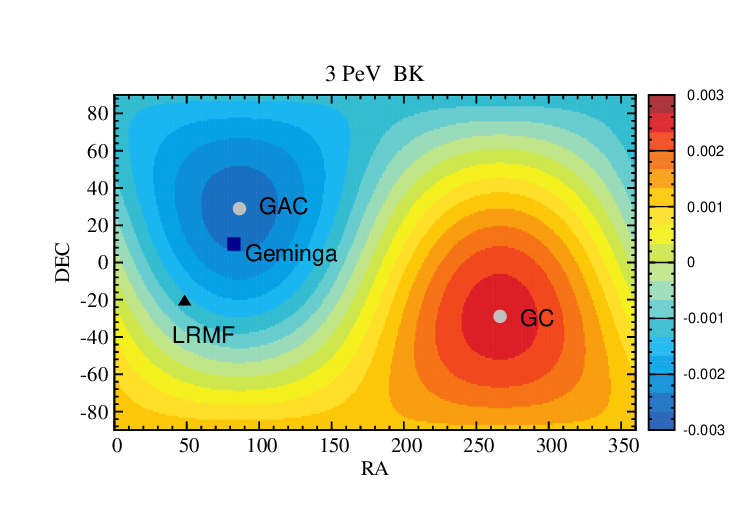}
	\includegraphics[width=0.45\textwidth]{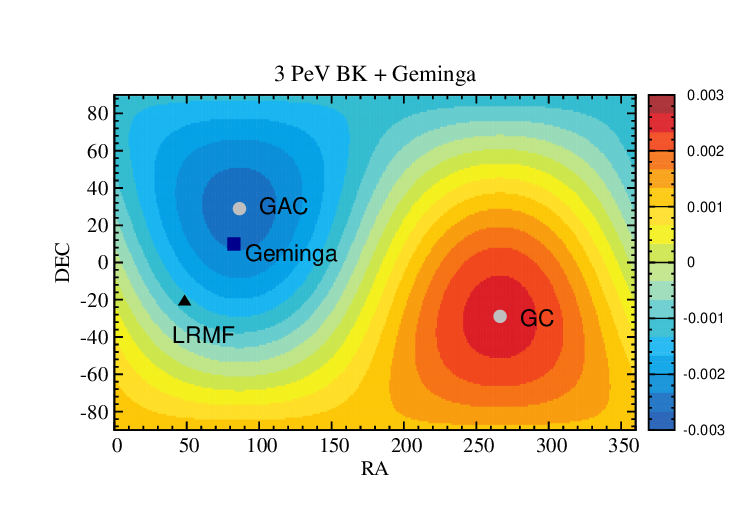}
	\includegraphics[width=0.45\textwidth]{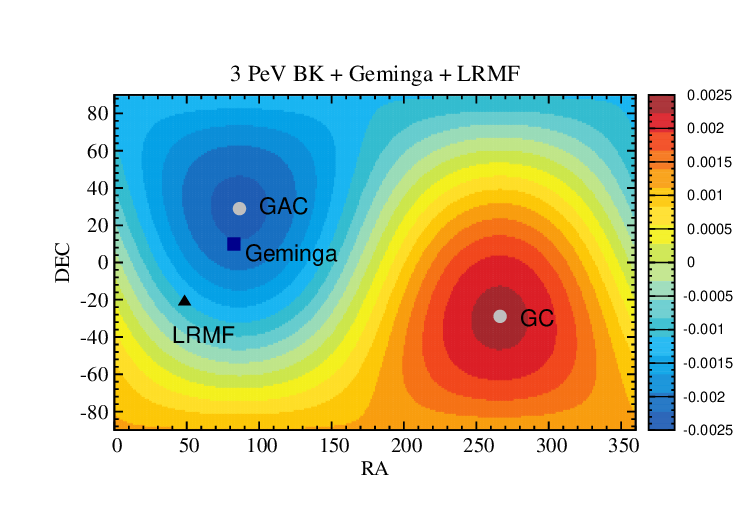}
	\end{adjustwidth}
\caption{Two-dimensional 	anisotropy maps at 10 TeV (up) and \mbox{3 PeV} (bottom), respectively, i.e. BK (left), BK + Geminga (middle), BK + Geminga + LRMF (right).
% Left maps are 2D anisotropy  under isotropic diffusion without contribution of nearby source (BK), middle maps are 2D anisotropy  under isotropic diffusion with the  contribution of nearby Geminga source (BK + Geminga), right mapa are  are 2D anisotropy  under anisotropic diffusion  introduced by LRMF with the  contribution of nearby Geminga source (BK + Geminga + LRMF).
}
	\label{fig:Fig2D}
\end{figure}

In order to further understand the effects of background sources, nearby source and LRMF on anisotropy in this model, 2D anisotropy sky maps with the contribution of each factor are presented. Figure \ref{fig:Fig2D} shows the 2D anisotropic sky map at 10 TeV (top) and 3 PeV (bottom), 
where the left, middle and right are the results of  background sources (BK), background sources + nearby Geminga source (BK + Geminga), and background sources + nearby Geminga  source +LRMF (BK + Geminga + LRMF), respectively.
It can be seen that in the low energy region, the anisotropy points to the GC without considering the contributions of nearby Geminga  source and LRMF, which is obviously contrary to the observation.
When the contribution of nearby Geminga source is introduced, the anisotropic phase points towards Geminga, which is attributed to  that the nearby source significantly alters the gradient of CR intensity in its direction. The results of BK + Geminga model are closer to the observed results, but there are still some deviations from the experimental observations. When the contribution of LRMF is further introduced, the anisotropy points in the direction of LRMF, which is completely consistent with the experimental results. This is due to the anisotropic diffusion effect of LRMF on CR particles.
However, in the high energy region, the results of anisotropy is relatively simple, and its phase always points to the GC. This indicates that the contribution of background sources is dominant, while the contribution of the nearby source is almost nonexistent in the high energy region.

\vspace{-12pt}

\section{Summary}\label{sec:summary}

%In our previous work, we have investigated GCR spectra and anisotropy from 100 GeV to PeV, under anisotropic propagation model with axisymmetric  distributed galactic sources. Numerous observations have indicated that the Milky Way possesses a spiral arm structure, which significantly influences the distribution and propagation of CRs.In this work, we further utilize anisotropic propagation models with  spiral galactic sources to investigate  energy spectra and anisotropy of CRs.During the calculation process, we employ the spatially dependent diffusion (SDP) model with different diffusion coefficients for the inner and outer halo, while the distribution of CR sources  are distributed  in a spiral pattern.
%while the distribution of background sources followed the spiral arm structure of the Galaxy.To better explain the anomalous observations of nuclear spectral hardening at$ {\cal R}\sim$ 200 GV and the complicated energy dependence of anisotropy from GeV to PeV, we introduce the contribution of the  Geminga nearby source. Additionally, we incorporate the impact of the local regular magnetic field (LRMF) and the corresponding anisotropic diffusion on large-scale anisotropy within the SDP model. From the results Our model successfully accounts for both CR spectra and anisotropy from 100 GeV to PeV. Future high-precision measurements of CR anisotropy, such as LHAASO experiment, will be crucial in evaluating the validity of our proposed model.

In recent years, a large number of scientific observations have demonstrated that the Milky Way possesses a spiral arm structure.  
Our previous work only analyzed the energy spectra and anisotropy based on the assumption of axisymmetric galactic source distribution.
The aim of this work is to explore the anisotropy and energy spectra by utilizing an anisotropic diffusion propagation model that incorporates a spiral distribution of background sources. 
Our model is based on the SDP model, while also accounting for  the contribution of  nearby Geminga source and the anisotropic diffusion effects of LRMF on CRs.
The results show that our model can simultaneously explain spectral hardening at 200 GeV and  the amplitude and phase variation of  anisotropy with energy from 100 GeV to PeV.
We also found that the  energy spectra and anisotropy  with spiral distribution  of  background  sources are similar to those with axisymmetric distribution of sources. Nevertheless, their propagation parameters are different. Specifically, the diffusion coefficient associated with the spiral distribution of sources is larger than that of the axisymmetric distribution, and the spectral indices for the spiral distribution are slightly harder.
This may be  attributed to the influence of the source distribution on the propagation of CRs.

We also studied the two-dimensional anisotropy sky maps that incorporate the contributions of nearby sources and LRMF.
Below the 100 TeV, it is clear that nearby Geminga  source contribute to the spectral hardening observed at 200 GeV.
Although the contribution of the nearby source to the CR flux is less significant compared to  that of the background sources, its impact on the anisotropy is dominant.
Under the isotropic diffusion model, the anisotropic phase is approximately oriented towards the nearby Geminga source. 
However, if the anisotropic diffusion effect of LRMF on CRs is taken into account, the anisotropic phase shifts to align with the direction of the LRMF.
In the high energy region above 100 TeV, the contribution of background sources becomes dominant, and the anisotropic phase consistently directs towards the GC.
Future measurements of CR spectra and anisotropies from higher-precision experiments, such as the LHAASO experiment, will provide valuable data to validate our model.
%Future CR spectrum and anisotropy measurements from more high-precision experiments, such as the LHAASO experiment, will help validate our model.

%\section*{Acknowledgements}

~\\
\textbf{Author Contributions:} Conceptualization, A.L. and W.L.; methodology, A.L. and Y.G.; software, F.Z.; validation, A.L., Y.G. and Z.L.; formal analysis, W.L.; investigation, A.L.; resources, A.L. and Z.L.; data curation, W.L.; writing—original draft preparation, A.L.and Y.G; writing—review and editing, A.L. and  W.L.; visualization, A.L. and Y.G.; supervision, A.L. and W.l.; project administration, W.L.; funding acquisition, A.L. All authors have read and agreed to the published version of the manuscript.

~\\
\textbf{Funding}: This work is supported by the National Natural Science Foundation of China (U2031110, 11963004, 12275279) and Shandong Province  Natural Science Foundation (ZR2020MA095). 

%%%%%%%%%%%%%%%%%%%%%%%%%%%%%%%%%%%%%%%%%%
\begin{adjustwidth}{-\extralength}{0cm}
	%\printendnotes[custom] % Un-comment to print a list of endnotes
	
	\reftitle{References}

\end{adjustwidth}

%%%%%%%%%%%%%%%%%%%%%%%%%%%%%%%%%%%%%%%%%%

%\bibliographystyle{unsrt_update}
%\bibliography{ref1}

%%%%%%%%%%%%%%%%%%%%%%%%%%%%%%%%%%%%%%%%%%

\end{document}